\documentclass[aps,twocolumn,prd,nofootinbib,showpacs]{revtex4}
\usepackage[pdftex,bookmarks]{hyperref}

\usepackage[T1]{fontenc}
\usepackage{amsmath,amssymb}

\usepackage{graphics,wrapfig,times}
\usepackage{graphicx}
\usepackage{color}
\usepackage{pause}
\usepackage{mathrsfs}

 \makeatletter
 \newcommand{\pback}[1]{{
   \let\@rrow=\leftarrowfill
   \mathchoice{\AIN@stemPullBack{#1}{\@rrow}}{\AIN@stemPullBack{#1}{\@rrow}}
     {\AIN@indxPullBack{#1}{\@rrow}}{\AIN@indxPullBack{#1}{\@rrow}}}
   \vphantom{#1}}

 \newcommand{\AIN@stemPullBack}[2]{
   \vtop{\mathsurround=0pt
   \ialign{##\crcr$\textstyle{#1}\strut$\crcr
     \noalign{\kern-0.4ex\nointerlineskip}{\tiny#2}\crcr}}}

 \newcommand{\AIN@indxPullBack}[2]{
   \vtop{\mathsurround=0pt
   \ialign{##\crcr\hfil$\scriptstyle{#1}$\hfil\crcr
     \noalign{\kern+0.4ex\nointerlineskip}{\tiny#2}\crcr}}}

\def\bar{\overline}
\def\be{\begin{equation}}
\def\ee{\end{equation}}
\def\bea{\begin{eqnarray}}
\def\eea{\end{eqnarray}}
\def\ba{\begin{array}}
\def\ea{\end{array}}
\def\nn{\nonumber}
\def\w{\wedge}
\def\s{\star}
\def\up{\stackrel}
\def\o{\textrm{out}}

\def\r{\mathrm{R}}
\def\t{\mathrm{T}}
\def\A{\mathcal{A}}
\def\C{\mathcal{C}}
\def\a{\alpha}
\def\b{\beta}
\def\c{\gamma}
\def\d{\delta}
\def\e{\epsilon}
\def\f{\kappa}
\def\g{\lambda}
\def\h{\mu}
\def\i{\nu}
\def\half{\frac{1}{2}}

\definecolor{D}{rgb}{0.00,0.17,0.48}
\definecolor{M}{rgb}{0.00,0.02,0.83}
\definecolor{L}{rgb}{0.58,0.79,1.00}
\definecolor{R}{rgb}{0.80,0.00,0.00}
\definecolor{G}{rgb}{0.02,0.40,0.10}

\begin{document}

\title{A momentum-space representation of Feynman propagator in Riemann-Cartan spacetime}

\author{Yu-Huei Wu} 
\email{yhwu@webmail.phy.ncu.edu.tw}
\affiliation{1. Center for Mathematics and Theoretical Physics, National Central University, \\
2. Department of Physics, National Central University, No. 300, Jhongda Rd., Jhongli 320, Taiwan
}

\author{Chih-Hung Wang} 
\email{chwang@phy.ncu.edu.tw, robbin1101@gmail.com}
\affiliation{ 1. Department of Physics, Tamkang University,
     Tamsui, Taipei 25137, Taiwan\\  2. Institute of Physics, Academia Sinica,Taipei 115, Taiwan 
    }

\begin{abstract}

We first construct generalized Riemann-normal coordinates by using autoparallels, instead of geodesics, in an arbitrary Riemann-Cartan spacetime. With the aid of generalized Riemann-normal coordinates and their associated orthonormal frames, we obtain a momentum-space representation of the Feynman propagator for scalar fields, which is a direct generalization of Bunch and Parker's works to curved spacetime with torsion. We further derive the proper-time representation in $n$ dimensional Riemann-Cartan spacetime from the momentum-space representation. It leads us to obtain the renormalization of one-loop effective Lagrangians of free scalar fields by using dimensional regularization. When torsion tensor vanishes, our resulting momentum-space representation returns to the standard Riemannian results.

\end{abstract}
\date{\today}
\maketitle


\section{Introduction}


A consistent dynamical theory between gravity and quantum field theory involves microscopic aspect of gravity. In the perturbative approach to quantize gravitational field, it turns out that general relativity (GR) is \textit{unrenormalizable}, so a satisfactory quantization of gravitational field cannot be accomplished \cite{Duff-75}. Moreover, the standard cosmological model, which is based on GR plus the known matter field, predicts a \textit{decelerating} expansion of our present Universe. It is contradicted to recent astrophysical observations, e.g. supernova Type Ia observations, which indicate that the expansion of the present Universe is in \textit{accelerating} phase \cite{Riess-et-al-98}. Hence, GR cannot successfully describe both the small-scale (planck scale) and large-scale (cosmological scale) phenomena. 

Since GR is established (by hypothesis) in the pseudo-Riemannian (i.e. torsion free) framework, the fundamental variables for the gravitational field are metric tensor $g$.  The sources of gravitational field are solely described by symmetric stress-energy tensor, so the conservation law of angular-momentum does not involve intrinsic spin of elementary particles. Therefore, it lacks of a description of spin-orbit coupling.  These problems may be resolved when we extend GR to Riemann-Cartan spacetime, i.e. Einstein-Cartan theory. Another well-known gauge theory of gravity, Poincar\'{e} gauge theory of gravity (PGT), is also established in Riemann-Cartan spacetime. In these theories, intrinsic spin play a significant role and becomes the source of torsion. 

Riemann-Cartan spacetime is characterized by metric $g$ and metric-compatible connection $\nabla$, so the natural associated variables for gravitational field are orthonormal co-frames $\{e^a\}$ and connection 1-forms $\{\omega^a{_b}\}$. In PGT, $\{e^a\}$ and $\{\omega^a{_b}\}$ correspond to local translation and rotation gauge potentials \cite{Hehl-76}.  There are several important theories of gravity being considered $\{e^a\}$ and $\{\omega^a{_b}\}$ as the fundamental variables, e.g. loop quantum gravity. Moreover, recent theoretical investigation on cosmology has been  developed in Riemann-Cartan spacetime.  For the dark energy problem,  it was discovered that the trace torsion may be a good candidate for dark energy in a cosmological model of PGT \cite{Shie-Yo-Nester-08, Chen-Ho-09}. For inflationary scenario in early Universe, instead of introducing inflaton which violate the strong energy condition, we found that the quadratic curvature terms in Riemann-Cartan spacetime do provide a power-law inflation and the totally anti-symmetric torsion play a significant role for generating inflation \cite{Wang-Wu-09}. These results show that torsion has notable effects on cosmology.

Since torsion and intrinsic spin have direct interactions, spin-polarized bodies are used to detect torsion directly in the laboratory (see a review article \cite{Ni-09}).  Up to present, there is no experimental evidences showing the existence of torsion field, so the constraints on torsion-spin coupling turn out to be extremely small \cite{Ni-09, KRT-08}. However, due to the observation of cosmic microwave background radiation (CMB) and other astrophysical observations, it provides another possibility to search for torsion-spin coupling, which are expected to be significant, in the early Universe. Instead of looking for torsion-spin coupling,  Dereli and Tucker considered a spinless particle following an \textit{autoparallel} curve in the Brans-Dicke theory with torsion, and then estimated the precession rate of Mercury's orbit \cite{Dereli-Tucker-01, Dereli-Tucker-02}.  Later on, the precession rate of a gyroscope following an autoparallel in the Kerr-Brans-Dicke field with torsion has also been calculated \cite{Wang-06}.     

The discovery of CMB and its anisotropic structure provides us a light to understand the evolution of our early Universe. It can be expected that the quantum effects of matter fields will become significant in the very early Universe (near the planck scale). In our previous work \cite{Wang-Wu-09}, we obtained an inflationary model based on quadratic curvature effects in Riemann-Cartan spacetime. In pseudo-Riemannian structure of spacetime, adding the quadratic curvature Lagrangians into the Einstein-Hibert action comes from the renormalization of one-loop effective action of matter fields \cite{Birrell-Davies} and inflationary models have also been discussed \cite{Starobinsky-80, MMS-86}. It motivates us to study the renormalization of one-loop effective action of matter fields in Riemann-Cartan spacetme.

Quantum field theory in the pseudo-Riemannian structure of spacetime has been largely investigated \cite{Birrell-Davies, DeWitt-65, DeWitt-75}. The standard approach to find the divergent terms of one-loop effective action of free scalar, spin 1/2 and 1 fields are using DeWitt-Schwinger proper-time method \cite{DeWitt-65, Christensen-78} with some regularization methods. It requires to solve a heat kernel equation in the \textit{normal} neighborhood of a point $x'$ by using DeWitt-Schwinger ansatz, which involve a bi-scalar world-function $\sigma (x, x')$, i.e. one-half the square of the geodesic distance between $x$ and $x'$. It is known that the normal neighborhood of $x'$ is obtained by using exponential map \cite{Hawking-Ellis}, i.e. sending geodesics to the neighborhood of $x'$. The generalization of proper-time formulation to Riemann-Cartan spacetime has been considered \cite{Goldthorpe-80, Nieh-Yan-82}. However, it immediately encounter with a question: which curve, autoparallel or geodeisc, should be used to construct the exponential map? In \cite{Goldthorpe-80}, it applied DeWitt-Schwinger ansatz to solve a heat kernel equation in Riemann-Cartan spacetime by using \textit{autoparallels}. However, we found that these curves are not autoparallels since the one-half the square of the \textit{autoparallel distance} $\sigma(x, x')$ satisfies the equation $\sigma(x, x') = \frac{1}{2} g^{\mu\nu}\nabla_{\mu}\sigma \nabla_\nu \sigma$ (see Eq. (3.8) in \cite{Goldthorpe-80}), which is actually the geodesic equation \cite{DeWitt-65}. It turns out that geodesic interval $\sigma(x, x')$ is more suitable for applying DeWitt-Schwinger ansatz in Riemann-Cartan spacetime. 

The discussion of quantum field theory in Riemann-Cartan spacetime has another approach by considering torsion as an extra background field (see a review article \cite{Shapiro-02}). In this approach, the fundamental variables are components of metric $g_{\mu\nu}$ and torsion $T^\alpha{_{\mu\nu}}$ with respect to coordinate basis $\{\partial_\mu \}$, and the full connection $\nabla$ will be separated into Levi-Civita connection $\tilde{\nabla}$ and contorsion part. Following this approach, the divergent terms of one-loop effective action of matter fields turn out to be the geometrical invariants associated with \textit{Riemannian} curvature and torsion, instead of full curvature and torsion \cite{Obukhov-83, Cognola-88, Shapiro-02}.  One may easily verify that the field equations obtained from the variation of $g_{\mu\nu}$ and $T^\alpha{_{\mu\nu}}$ are completely different from the field equations obtained by varying $\{e^a\}$ and $\{\omega^a{_b}\}$. For example, the metric components $g_{\mu\nu}$ contain 10 independent variables due to its symmetrization, but orthonormal-coframe $\{e^a= e^a{_\mu} d x^\mu\}$ have 16 independent variables. So their associated source currents will be \textit{symmetric} stress-energy tensor $T^{\mu\nu}$ and stress-energy 3-forms $\tau_a$, respectively. Moreover, another difference comes from the first-order and second-order Lagrangians \cite{Shapiro-02}. It is more reasonable for us to consider $\{e^a\}$ and $\{\omega^a{_b}\}$ as independent variables, since one can naturally derive their associated field strengths (i.e. torsion and full curvature) and Bianchi identities (i.e. conservation laws).  

Besides the DeWitt-Schwinger proper-time representation, Bunch and Parker \cite{Bunch-Parker-79} developed a momentum-space representation which is useful for discussing the renormalizability of interacting fields, e.g. $\lambda \phi^4$ theory, in a general pseudo-Riemannian structure of spacetime. By constructing the Riemann-normal coordinates in the normal neighborhood of an original point $x'$, they solved the Feynamm Green's function $G (x, x')$ of free scalar and Dirac fields in the Riemann-normal coordinates with a large wave number $k$ approximation. In the large $k$ approximation, the first few leading solution of momentum-space representation of Feynamm Green's function, i.e. $\bar{G}_i(k)$, have been obtained. It is known that the ultraviolet divergences come from the large wave number $k$ modes, i.e. short distance behavior, so the solutions $\bar{G}_i(k)$ for $i\leqslant 4$ are sufficient to deal with the renormalization of one-loop effective action. However, for studying the renormalizability of $\lambda \phi^4$ theory, the solutions $\bar{G}_i (k)$ for $i \leqslant 2$ may be sufficient.  

The method of momentum-space representation can be naturally extended to Riemann-Cartan spacetime. The major different is that the background field variables are changed from metric tensor $g= g_{\mu\nu} d x^\mu \otimes d  x^\nu$ to orthonormal co-frames $\{e^a=e^a{_\mu} d x^\mu\}$ and connection 1-forms $\{\omega^a{_b}= \omega^a{_{b\mu}} d x^\mu \}$. We should construct a local coordinate system $\{x^\mu\}$, where the coefficients of $e^a{_\mu}$ and $\omega^a{_{b\mu}}$ in the Taylor series expansions can be systematically expressed in terms of full curvature, torsion and their covariant derivative $\nabla_\mu$ at original point $x'$.  It is obvious that the Riemann-normal coordinate is not a proper choice since only the expansions of $g_{\mu\nu}$ can be systematically expressed in terms of \textit{Riemannian} curvature and its covariant derivative with respect to Levi-Civita connection. There is no systematical way to accomplish the expansions of full connection components $\Gamma^\a{_{\mu\nu}}$. It is not difficult to see that $g_{\mu\nu}$ in the Riemann-normal coordinates yields \textit{no} difference in the Riemann-Cartan spacetime or in the pseudo-Riemannian structure of spacetime. 

Tucker established Fermi coordinates with their associated orthonormal-frames in Riemann-Cartan spacetime \cite{Tucker-04}. Instead of using geodesics, he used autoparallels $\gamma_v(\lambda)$ to define an exponential map and then the Fermi coordinates can be constructed in the normal neighborhood of a time-like curve. By parallel transporting the orthonormal co-frames $\{e^a\}$ along $\gamma_v(\lambda)$, one can systematically expressed $e^a{_\mu}$ and $\omega^a{_{b\mu}}$ in terms of the acceleration and frame rotation of the time-like curve, full curvature, torsion and $\nabla_\mu$ on the time-like curve. We follow the similar process to construct generalized Riemann-normal coordinates at a point $x'$. A detail construction will be presented in Sec. \ref{3}. A recent investigation on normal frames in general connection (no metric-compatible) has be found in \cite{Nester-09}. In order to find the ultraviolet divergences of one-loop effective action, we should accomplish the expressions of $e^a{_\mu}$ and $\omega^a{_{b\mu}}$ in terms of full curvature, torsion and $\nabla_\mu$ to fifth-order, which involve quadratic full curvature terms. 

Using generalized Riemann-normal coordinates, one can solve Feynamm Green's function of scalar and spin 1/2 fields in Riemann-Cartan spacetime and find approximate solutions $\bar{G}_i (k)$ in the momentum space. In this paper, we first concentrate on a scalar field. The classical action of the scalar field includes a non-minimal coupling term $\xi R \phi^2$, where $R$ is full scalar curvature. When $\xi=0$, the scalar field has miminal coupling, and $\xi= \frac{1}{6}$  may refer to conformal coupling. Since solving $\bar{G}_i(k)$ for $i=3, 4$ involves extremely complicated and tedious calculations, we will restrict our background torsion to be \textit{totally} ansi-symmetric. For $i\leqslant 2$, we solve $G_i(k)$ in general background torsion. The restriction of totally anti-symmetric torsion largely simplified our calculations. Moreover, our previous work \cite{Wang-Wu-09} indicated that totally anti-symmetric torsion plays a significant role for generating inflation, so this restriction may still be useful for investigating quantum effects in the early Universe. The calculation of renormalization of one-loop effective action for spin 1/2 field is straightforward and since the totally anti-symmetric torsion has a direct interaction with fermions, the restriction on totally anti-symmetric torsion may also be interesting to study.

In Sec. \ref{2}, we start from a classical action of a free massive scalar field in Riemann-Cartan spacetime, and by using path-integral quantization,  the effective action is obtained. The vacuum expectation values of stress 3-forms and spin 3-forms are defined. Sec. \ref{3} presents a detail construction of generalized Riemann-normal coordinates with associated orthonormal frames. We derive the expansions of  $e^a{_\mu}$ and $\omega^a{_{b\mu}}$ to fifth-oder, and the coefficients are expressed in terms of full curvature, torsion and $\nabla_\mu$ at the original point $x'$. When torsion vanishes, it agrees with the result obtained in the Riemann-normal coordinates.  Sec. \ref{4} starts from the equation of Feynamm Green's function in $n$ dimensional Riemann-Cartan spacetime, and by using the generalized Riemann-normal coordinates constructed in Sec. \ref{3}, and large $k$ approximation, we obtain the solutions $\bar{G}_i(k)$ for $i\leqslant 2$ in general background torsion. In the Subsection \ref{4-1}, the solutions $\bar{G}_i (k)$ for $i \leqslant 4$ are derived in the totally antisymmetric background torsion. When torsion vanishes, $\bar{G}_i(k)$ agrees with the result in \cite{Bunch-Parker-79}. In Sec. \ref{5}, the proper-time representation in $n$ dimensional Riemann-Cartan spacetime is derived from the momentum-space representation obtained in Sec. \ref{4}. Since the solutions $\bar{G}_i(k)$ are valid in $n$ dimensional spacetime, we use dimensional regularization to study the renormalization of one-loop effective action. In Appendix, we present the detail and tedious calculations for writing down the equation of Feymann Green's function in the generalized Riemann-normal coordinates.

In this paper, we use unit $h=c=1$, and for $n$ dimensional spacetime, the metric signature is $(-, +, \cdots, +)$.  The Greek indices $\a,\b,\c \cdots$ are referred to coordinate indices and the Latin indices $a, b, \cdots$ referred to frame indices. Both types of indices run from $0$ to $n-1$. The covariant derivative $\nabla_\h$ on any tensor components $Z^{a\cdots b}{_{c\cdots d}}$ is defined by $(\nabla Z)( e^a,\cdots, e^b, X_c, \cdots, X_d, \partial_\h)$. Any geometrical object defined by Levi-Civita connection $\tilde{\nabla}$ will be put $\tilde{ }\,$ on it.


\section{Effective Lagrangians of scalar fields} \label{2}

 The classical action functional of a scalar field in the pseudo-Riemannian (i.e. torsion free) structure of space-time is \cite{Birrell-Davies}
\bea
\tilde{S} [ g, \phi ] = - \frac{1}{2} \int_M d \phi \w \s d \phi + ( m^2 + \xi \tilde{R}) \phi^2 \star 1, \label{as-1}
\eea
 where the metric $g$ denotes the background gravitational field, $\star$ is the Hodge map associated with $g$, $m$ is the scalar field's mass, $\xi$ is an arbitrary real number, and $\tilde{R}$ is the Ricci scalar curvature defined by Levi-Civita connection $\tilde{\nabla}$.  Since the background gravitational field is now described by $g$ and metric-compatible connection $\nabla$ in the Riemann-Cartan space-time, the basic gravitational variables will be a class of arbitrary local orthonormal 1-form co-frames $\{e^a\}$ on space-time related by $SO( 3, 1)$ transformation and connection 1-forms $\{\omega^a{_b}\}$, which is a representation of $\nabla$ with respect to $\{e^a\}$.  A direct generalization of Eq. (\ref{as-1}) to Riemann-Cartan space-time is 
\bea
S [e^a, \omega^a{_b}, \phi] = - \frac{1}{2} \int_M d \phi \w \s d \phi + ( m^2 + \xi R ) \phi^2 \s 1, \label{as-2}
\eea where $R$ is the full scalar curvature. Varying $S$ with respect to $\phi$, the equations of motion of $\phi$ can be obtained
\bea
0=\frac{\delta S}{\delta \phi} = - d \s d \phi + ( m^2 + \xi R ) \phi \s 1. \label{seq-1}
\eea The classical stress 3-forms $\tau_a$ and spin 3-forms $S_a{^b}$ are defined as
\bea
\tau_a &\equiv& \frac{\delta S}{\delta e^a} =\frac{1}{2}( i_a d \phi \w \s d \phi + d \phi \w i_a \s d \phi - m^2 \phi^2 \s e_a \nn\\
 && \hspace{1.5cm}- \xi \phi^2 \,\r_{bc}\w \s e_{a}{^{bc}} )  ,  \\
 S_a{^b}&\equiv& \frac{\delta S}{\delta \omega^a{_b}} = - \frac{1}{2}\, \xi \phi^2 \left( \frac{2\,d \phi}{\phi} \w \s e_a{^b} +\t^c \w \s e_{ca}{^b} \right), 
\eea where $i_a \equiv i_{X_a}$ is the interior derivative and $\{X_a\}$ is the dual basis of $\{e^a\}$. $e^{a\ldots b}{_{c\ldots d}} \equiv e^a \w \ldots \w e^b \w e_c \w \ldots e_d$, $\r_{ab}$ are curvature 2-forms, and $\t^a$ are torsion 2-forms.

The transition from classical to quantum fields has two main procedures, canonical and path-integral quantizations.  However,  the path-integral quantization is a more practical approach to study the renormalization of vacuum expectation value of $\tau_a$ and $S_a{^b}$,
which are given by 
\bea
&&<\tau_a> = \frac{<\o, 0| \tau_a| \textrm{in}, 0>}{<\o, 0| \textrm{in}, 0>}, \label{tau}\\
&&<S_a{^b}>=  \frac{<\o, 0| S_a{^b}| \textrm{in}, 0>}{<\o, 0|\textrm{in} , 0>} \label{S},
\eea where $|\textrm{in}, 0>$ and $|\o, 0>$ correspond to initial in-region and final out-region vacuum states, respectively. 

One may start from the generating functional with vanishing external current $J=0$ \cite{Birrell-Davies},
\bea
Z[J=0] \equiv <\o, 0| \textrm{in}, 0> = \int \mathcal{D} [\phi] \,e^{i S[e^a, \omega^a{_b}, \phi]},
\eea and using Schwinger's variational principle \cite{DeWitt-65} yields
\bea
\delta Z[0] = i <\o, 0| \delta S| \textrm{in}, 0>= i \int \mathcal{D} [\phi] \, \delta S \,e^{i S}.
\eea So $<\tau_a>$ and $<S_a{^b}>$ can actually be derived by varying the effective action $W$, which is defined by  
\bea
W[e^a, \omega^a{_b}, \phi ] \equiv - i \ln <\o, 0| \textrm{in}, 0> = - i \ln Z[0], \label{W}
\eea  with respect to $e^a$ and $\omega^a{_b}$, i.e., 
\bea
&&\frac{\delta W}{\delta e^a} = <\tau_a>, \\
&&\frac{\delta W}{\delta \omega^a{_b}} = <S_a{^b}>.
\eea 

With some straightforward derivation \cite{Birrell-Davies}, the effective action become 
\bea
W= - \frac{i}{2}\int_M <x | \ln G_F | x > \star 1, \label{EA}
\eea where $<x | G_F | x^\prime> \equiv G_F(x, x^\prime)$ is the Feynman Green's function. The divergent terms of $W$ have been largely studied by using the Dewitt-Schwinger proper-time representation of $G_F(x, x^\prime)$ in the pseudo-Riemannian geometry  \cite{Birrell-Davies, DeWitt-65}. Besides the proper-time representation, Bunch and Parker proposed another representation, the momentum-space representation, which has been shown to be equivalent to proper-time representation , to study the renormalizability of $\lambda \phi^4$ field theory in the pseudo-Riemannian geometry \cite{Bunch-Parker-79}. 

The momentum-space representation of $G_F(x, x^\prime)$ requires to establish the Riemann-normal coordiantes in a normal neighborhood of the point $x^\prime$ and then solve $G_F(x, x^\prime)$ in the momentum space \cite{Bunch-Parker-79}. The resulting divergent terms of effective action involve Ricci scalar curvature $\tilde{R}$ and various quadratic Riemann curvature terms, e.g. Ricci curvature square $\tilde{R}_{ab} \tilde{R}^{ab}$.  The method of  momentum-space representation can naturally be extended to Riemann-Cartan spacetime, however, it is not suitable to use the Riemann-noraml coordinates. The reason is that divergent terms of Eq. (\ref{W}) should depend on \textit{full} curvature and torsion terms instead of Riemann curvature and torsion. It leads us to establish a generalized Riemann-normal coordinates in a Riemann-Cartan spacetime.

\section{Generalized Riemann-normal coordinates} \label{3}

In a general Riemann-Cartan spacetime, the definitions of autoparallels and geodesics are completely different.  However, they become equivalent in the pseudo-Riemannian geometry. Autoparallels $\gamma : \lambda \mapsto \gamma(\lambda)$, which satisfy
\bea
\nabla_{\gamma^\prime} \gamma^\prime =0, \label{ap}
\eea where $\gamma^\prime$ denotes the tangent vector of $\gamma$, are defined in terms of connectiton $\nabla$, but geodesics $C : t \mapsto C(t)$, which satisfy 
\bea
\delta \int \sqrt{g ( \dot{C}, \dot{C})}\,\, d t =0, \label{ge}
\eea are defined in terms of $g$. It is worth to point out that autoparallels and geodesics coincide if background torsion tensor requires to be totally anti-symmetric.  Since Eq. (\ref{ap}) and Eq. (\ref{ge}) both provide unique solutions with given initial values, it is not difficult to see that either autoparallels or geodesics can be used to construct a local coordinate system.  

If one uses geodesics to construct local coordinates $y^\alpha$, i.e., the Riemann-normal coordinates, with original at point $x^\prime$ in the Riemann-Cartan spacetime, the expansion of the metric components $g_{\mu\nu}$ (i.e. $g(\partial_\mu, \partial_\nu)$) in this coordinates is given by \cite{Petrov-69}
\bea
g_{\mu\nu} = \delta_{\mu\nu} - \frac{1}{3} \tilde{R}_{\mu\nu\alpha\beta}y^\alpha y^\beta - \frac{1}{6} \tilde{\nabla}_\gamma \tilde{R}_{\mu\nu\alpha\beta} y^\alpha y^\beta y^\gamma +\ldots, \nn\\
\eea which involves the value of the Riemann curvature $\tilde{R}_{abcd}$ and the covariant derivative with respect to Levi-Civita connection $\tilde{\nabla}$ at the original point, instead of $full$ curvature $R_{abcd}$ and connection $\nabla$, at $x^\prime$. Furthermore, it can be verified that the expansions of all geometric quantities, e.g. torsion tensor components $T^a{_{bc}}$,  also involve $\tilde{R}_{abcd}$ and $\tilde{\nabla}$.   Since our background gravitational variables are $\{e^a\}$ and $\{\omega^a{_b}\}$, using the above construction to find the divergent terms of $W$ is completely improper. It is necessary to establish a local coordinate system, where the expansions of $\{e^a\}$ and $\{\omega^a{_b}\}$ will involve full curvature $R^a{_{bcd}}$, covariant derivative $\nabla_a$, and torsion $T^a{_{bc}}$. The generalized Fermi coordinates have been constructed by using autoparallels and the associated orthonormal co-frames in the Riemann-Cartan spacetime \cite{Tucker-04}. Here, we apply the similar procedure to establish generalize Riemann-normal coordinates.

Consider an autoparallel $\gamma_v : \lambda\mapsto\gamma_v(\lambda) \in M$ with its initial values
\bea
&&\gamma_v (0) = x^\prime, \\
&&\gamma^\prime_v (0) =v, 
\eea where $M$ denotes a $n$-dimensional Riemann-Cartan spcaetime. Provided $\gamma_v (1)$ exists, the exponential map $\exp_{x^\prime} : T_{x^\prime}M \mapsto M$ is then defined in an open neighborhood $\mathcal{U}$ of $x^\prime$  by
\bea
\exp_{x^\prime} (v) \equiv \gamma_v (1) \in M,
\eea where $T_{x^\prime}M$ denotes the tangent space to $M$ at $x^\prime$. Using $\exp_{x^\prime}$ with an orthonormal frame $\{\hat{X}_a\}$ at $x^\prime$,  we obtain the generalized Riemann-normal coordiantes $x^\a$
\bea
\Psi^\a (\exp_{x^\prime} v ) = x^\a,
\eea where $\Psi^\a$ is a coordinate chart and
\bea
v=\sum^{n-1}_{\a=0} \delta^a{_\a}\, x^\a \hat{X}_a,
\eea where $\delta^a{_\a} = \textrm{diag}(1, \cdots, 1)$. In the following,  $\hat{ }\,$ on any tensor field $Z$ denotes $Z|_{x^\a =0}$ (i.e. $Z$ at $x^\prime$). A natural induced coordinate basis $\{\partial_\a\}$, by construction, has $\{\hat{\partial}_\a = \delta^a{_\a}\hat{X}_a\}$.

It will be useful to introduce generalized Riemann-normal hyper-spherical coordinates $\{\lambda, \,p^\a\}$ defined by 
\bea
x^\a = \lambda\, p^\a,
\eea where $\lambda$ is the radial coordinate with affine parametrization and $p^a$ are the direction cosines of tangent vectors of autoparallels  $\gamma_{\partial_\a}$ at $x^\prime$ satisfying 
\bea
\sum^{n-1}_{\a=0} p^\a p^\a = 1.
\eea  From the inverse relations
\bea
\lambda^2 = \sum^{n-1}_{\a=0} x^\a x^\a, 
\eea one has 
\bea
&&\partial_{\lambda} = p^\a \partial_{\a}, \\
&&\partial_{\lambda} p^\a =0,
\eea and $v= \lambda \hat{\partial}_\lambda$. It should be mentioned that $\hat{Z}= Z|_{\lambda=0}$ denotes the initial value of any tensor field $Z$ in hyper-spherical coordinates $\{\lambda, \, p^\a\}$.  Using $\{\lambda,\, p^\a\}$, we can parallel transport $\{\hat{X}_a\}$ along autoparallels $\gamma_{\partial_\lambda}$ to set up a field of orthonormal frames $\{X_a\}$ and its dual co-frame field  $\{e^a\}$ on $\mathcal{U}$.  

From the above construction, one has
\bea
\nabla_{\partial_\lambda} e^a = 0, \label{e^a}
\eea i.e.
\bea
i_{\partial_\lambda} \omega^a{_b}= \omega^a{_b} (\partial_\lambda) =0, \label{omega}
\eea with its initial value $\hat{e}^a = \delta^a{_\a}\, \widehat{d x^\a} = \delta^a{_\a}\, p^\a \widehat{d \lambda}$. Since $\partial_\lambda$ are tangent vectors of autoaparallels, we further obtain 
\bea
\partial_\lambda (e^a(\partial_\lambda)) =0.
\eea It turns out that  $e^a (\partial_\lambda)$ is independent of $\lambda$ and equals to its intital value $p^a$. So $\{e^a\}$ in $\{\lambda, p^a\}$ gives
\bea
e^a = \delta^a{_\a}\, p^\a d \lambda + \mathcal{A}^a{_\h}\, d p^\h \label{e^a-1}
\eea with the initial values 
\bea
\hat{\mathcal{A}}^a \equiv \hat{\mathcal{A}}^a{_\h}\, d p^\h = 0. \label{initial-1}
\eea  Eq. (\ref{omega}) indicates that $\omega^a{_b}$ do not contain the $d\lambda$ term, so
\bea
\omega^a{_b} = \mathcal{C}^a{_{b\h}}\, d p^\h \label{omega-1}
\eea with the initial values
\bea
\hat{\mathcal{C}}^a{_b}=\hat{\mathcal{C}}^a{_{b\h}}\, d p^\h = \widehat{\omega^a{_b} (\partial_\h)} \,d p^\h =0. \label{initial-2}
\eea 

It is known that the Riemann-normal coordinates in the pseudo-Riemannian geometry have a local Minkwoski structure (i.e. $g_{\h\i}(x^\prime)= \tilde{\omega}^\h{_\i}(x^\prime) =0$), which is associated with equivalence principle. Similarly,  Eqs. (\ref{initial-1}) and (\ref{initial-2}) also  represent a local Minkwoski structure of spacetime at $x^\prime$ in the Riemann-Cartan spacetime, so the revised edition of equivalence principle has been discussed \cite{Heyde-75}.

Since we have completely constructed the generalized Rieamnn-normal coordinates with the associated orthonormal co-frames $\{e^a\}$ on $\mathcal{U}$, the next step is to expand the fundamental variables $\{e^a\}$ and $\{\omega^a{_b}\}$ with respect to radial variable $\lambda$ and then to express their coefficients in terms of full curvature $\hat{R}^a{_{bcd}}$, torsion $\hat{T}^a{_{bc}}$ and their covariant derivative $\nabla_\a$. 

We start from the Cartan structure equations defined by the torsion and full curvature \cite{BT-87}:
\bea
&&d e^a = - \omega^a{_b} \wedge e^b + \t^a, \label{se-1}\\
&& d \omega^a{_b} = - \omega^a{_c} \wedge \omega^c{_b} + \r^a{_b} \label{se-2}
\eea where
\bea
\t^a = \frac{1}{2} T^a{_{bc}} \,e^{bc}\hspace{0.5cm} \textrm{and}\hspace{0.5cm}
\r^a{_b} =  \frac{1}{2} R^a{_{bcd}}\, e^{cd}
\eea are torsion 2-forms and curvature 2-forms in the co-frame field $\{e^a\}$. By substituting Eqs. (\ref{e^a-1}) and (\ref{omega-1}) into Eqs. (\ref{se-1}) and (\ref{se-2}) and equating the forms containing $d\lambda \w d p^\a$ on each side, it gives ordinary differential equations for $\mathcal{A}^a$ and $\mathcal{C}^a{_b}$:
\bea
&&\mathcal{A}^{\prime a} = \delta^a{_\a}\,d p^\a + \mathcal{C}^a{_b}\,\delta^b{_\a}\, p^\a + T^a{_{bc}}\,\delta^b{_\a}\, p^\a \mathcal{A}^c, \label{sc-3}\\
&&\mathcal{C}^{\prime a}{_b} = R^a{_{bcd}}\,\delta^c{_\a}\, p^\a \mathcal{A}^d, \label{sc-4}
\eea where $\prime$ denotes the radial derivative $\partial_\lambda$. $\A'^a$ and $\C'^a{_b}$ denote $(\partial_\lambda \mathcal{A}^a{_b}) \, d p^b$ and $\C'^a{_b} = (\partial_\lambda \mathcal{C}^a{_{bc}}) \, d p^c$, respectively. In the remain part of Sec. \ref{3}, we will use the notations $d p^a \equiv \delta^a{_\a}\, d p^\a$ and $p^a \equiv \delta^a{_\a}\, p^\a$

We know that the Taylor series representations of $\A^a$ and $\C^a{_b}$ with respect to radial coordinate $\lambda$ are
\bea
&&\A^a = \hat{\A}^a + \hat{\A}'^a \lambda + \frac{1}{2 !} \hat{\A}''^a \lambda^2+
\cdots, \\
&&\C^a{_b} = \hat{\C}^a{_b} + \hat{\C}'^a{_b} \lambda + \frac{1}{2 !} \hat{\C}''^a{_b} \lambda^2+
\cdots.
\eea It should be mentioned that, for any function $f$, $\hat{f}^{\prime \cdots \prime}$ denotes $(\partial_\lambda \cdots \partial_\lambda f )|_{\lambda=0}$.
By successively differentiating Eq. (\ref{sc-3}) and (\ref{sc-4}) with respect to $\lambda$ and then evaluating the results at $\lambda=0$, one can obtain $\hat{\mathcal{A}}^{\prime \cdots \prime a}$ and $\hat{\mathcal{C}}^{\prime \cdots \prime a}{_b}$ in terms of $\hat{R}^a{_{bcd}}$, $\hat{T}^a{_{bc}}$, and their radial derivative $\partial_\lambda$. Since the discussion of renormalization of $W$ in terms of momentum-space representation requires to calculate $\hat{\mathcal{A}}^{\prime \cdots \prime a}$ and $\hat{\mathcal{C}}^{\prime \cdots \prime a}{_b}$ to \textit{fifth}-order, we will present our results to fifth-order of the radial derivative. To first order in $\lambda$ one find
\bea 
&&\hat{\A}'^a = d p^a, \label{e1}\\ 
&&\hat{\C}'^a{_b} =0.  
\eea The curvature and torsion start to appear at the second order:
\bea 
&&\hat{\A}''^a =  \hat{T}^a{_{bc}}\, p^b \,d p^c, \label{e2}\\
&&\hat{\C}''^a{_b} = \hat{R}^a{_{bcd}}\, p^c \,d p^d.
\eea At the third order:
 \bea 
 \hat{\A}'''^a &=& \hat{R}^a{_{bcd}}\, p^b p^c d p^d + 2 \hat{T}'^a{_
{bc}}\, p^b d p^c \nn\\
&&+ \,\hat{T}^a{_{bc}}\, \hat{T}^c{_{de}}\, p^b p^d d p^e, \label{e3}\\
 \hat{\C}'''^a{_b} &=& 2 \hat{R}'^a{_{bcd}}\, p^c d p^d + \hat{R}^a{_
{bcd}}\, \hat{T}^d{_{ef}}\,p^c p^e d p^f,
\eea which have one radial derivative of the curvature and torsion. 
\begin{widetext}
The two radial derivatives of the curvature and torsion start to appear at the fourth order:
 \bea 
 \hat{\A}''''^a &=& 2\hat{R}'^a{_{bcd}}\, p^b p^c d p^d +  \hat{R}'^a{_
{bcd}}\,\hat{T}^d{_{ef}}\, p^b p^c p^e d p^f 
+3 \hat{T}''^a{_{bc}}\, p^b d p^c + \hat{T}^a{_{bc}}\, \hat{R}^c\,_{def}\, p^b p^d p^e  d p^f \nn\\
 &&+ 3 \hat{T}'^a{_{bc}}\, \hat{T}^c{_{de}}\, p^b
p^d d p^e +2 \hat{T}^a{_
{bc}} \,\hat{T}'^c{_{de}}\, p^b p^d d p^e 
 + \hat{T}^a{_{bc}} \, \hat{T}^c{_{de}} \, \hat{T}^e{_{fg}}\,p^b p^d p^f d p^g,  \label{e4}\\ 
 \hat{\C}''''^a\,_b &=& 3 \hat{R}''^a{_{bcd}}\, p^c d p^d + \hat{R}^a{_{bcd}} \, \hat{R}^d{_{efg}}\, p^c p^e p^f d p^g  
 + 3\hat{R}'^a{_
{bcd}} \, \hat{T}^d{_{ef}} \,p^c p^e d p^f + 2 \hat{R}^a{_
{bcd}} \, \hat{T}'^d{_{ef}}\,p^c p^e d p^f\nn\\
 &&+ \hat{R}^a{_{bcd}}\,  \hat{T}^d{_{ef}} \, \hat{T}^f{_{gh}}\,p^c p^e p^g d p^h.
\eea
 At the fifth order, it becomes much more complicated and involves three radial derivatives of the curvature and torsion:
 \bea 
 \hat{\A}'''''^a &=& 3 \hat{R}''^a{_{bcd}}\, p^b p^c d p^d + 3 \hat{R}'^a{_{bcd}} \, \hat{T}^d{_{ef}} \,p^b p^c p^e d p^f
       +\hat{R}^a{_{bcd}}\, \hat{R}^d{_{efg}}\,p^b p^c p^e p^f d p^g + 2\hat{R}^a{_{bcd}}\, \hat{T}'^d{_{ef}}\, p^b p^c p^e d p^f \nn\\
       &+& \hat{R}^a{_{bcd}} \, \hat{T}^d{_{ef}} \, \hat{T}^f{_{gh}}\,p^b p^c p^e p^g d p^h
       + 4\hat{T}'''^a{_{bc}}\, p^b d p^c + 6 \hat{T}''^a{_{bc}} \, \hat{T}^c{_{ed}}\,p^b p^e d p^d
       + 4\hat{T}'^a{_{bc}} \, \hat{R}^c{_{def}}\,p^b p^d p^e d p^f \nn\\
       &+& 8\hat{T}'^a{_{bc}}\, \hat{T}'^c{_{de}}\, p^b p^d d p^e
       + 4\hat{T}'^a{_{bc}} \, \hat{T}^c{_{de}} \, \hat{T}^e{_{fg}}\,p^b p^d p^f d p^g
       + 2\hat{T}^a{_{bc}}\,  \hat{R}'^c{_{def}}\,p^b  p^d p^e d p^f \nn\\
       &+&\hat{T}^a{_{bc}}\, \hat{R}^c{_{def}}\, \hat{T}^f{_{gh}}\,p^b p^d p^e p^g d p^h
       + 3\hat{T}^a{_{bc}}\, \hat{T}''^c{_{de}} \, p^b p^d d p^e + 3\hat{T}^a{_{bc}}\, \hat{T}'^c{_{de}}\, \hat{T}^e{_{fg}}\,p^b  p^d p^f dp^g\nn\\
       &+& \hat{T}^a{_{bc}} \, \hat{T}^c{_{de}}\,  \hat{R}^e{_{fgh}} \,p^b p^d p^f p^g d p^h
       + 2\hat{T}^a{_{bc}}\, \hat{T}^c{_{de}} \, \hat{T}'^e{_{fg}} \,p^b p^d p^f d p^g
       + \hat{T}^a{_{bc}}\,  \hat{T}^c{_{de}} \, \hat{T}^e{_{fg}}\, \hat{T}^g{_{hi}} \, p^b  p^f p^d p^h d p^i \label{e5} \\
 %
 \hat{\C}'''''^a{_b} &=& 4 \hat{R}'''^a{_{bcd}}\, p^c d p^d + 6 \hat{R}''^a{_{bcd}}\, \hat{T}^d{_{ef}}\, p^c p^e d p^f
 + 4 \hat{R}'^a{_{bcd}}\, \hat{R}^d{_{efg}}\, p^c p^e p^f d p^g +  8 \hat{R}'^a{_{bcd}}\, \hat{T}'^d{_{ef}}\,p^c p^e d p^f\nn\\
 &+& 4 \hat{R}'^a{_{bcd}}\, \hat{T}^d{_{ef}} \, \hat{T}^f{_{gh}}\,p^c p^e p^g d p^h
 + 2 \hat{R}^a{_{bcd}}\, \hat{R}'^d{_{efg}}\,p^c p^e p^f d p^g + \hat{R}^a{_{bcd}}\, \hat{R}^d{_{efg}} \, \hat{T}^g{_{hi}}\, p^c p^e p^f p^h d p^i \nn\\
 &+& 3 \hat{R}^a{_{bcd}}\,  \hat{T}''^d{_{ef}} \, p^c p^e d p^f  + 3 \hat{R}^a{_{bcd}}\, \hat{T}'^d{_{ef}} \, \hat{T}^f{_{gh}}\,p^c p^e p^g d p^h
 +   \hat{R}^a{_{bcd}}\, \hat{T}^d{_{ef}} \, \hat{R}^f{_{ghi}}\,p^c p^e p^g p^h d p^i\nn\\
 &+& 2 \hat{R}^a{_{bcd}}\,  \hat{T}^d{_{ef}} \, \hat{T}'^f{_{gh}}\,p^c p^e p^g d p^i
 + \hat{R}^a{_{bcd}} \, \hat{T}^d{_{ef}} \, \hat{T}^f{_{gh}} \, \hat{T}^h{_{ij}}\,p^c p^e p^g p^i d p^j  \eea

Although these expressions involve the radial derivative $\partial_\lambda$, it can be changed to covariant derivative $\nabla_{\partial_\lambda}$ by using Eq. (\ref{e^a}), e.g. 
\bea
\nabla_{\partial_\lambda} R^a{_{bcd}} \equiv (\nabla_{\partial_\lambda} R) (e^a, X_b, X_c, X_d) 
= \nabla_{\partial_\lambda} \left( R (e^a, X_b, X_c, X_d) \right) \equiv \partial_\lambda R^a{_{bcd}}.
\eea Moreover, it is understood that any tensor-field components $Z^{a\cdots b}{_{c\dots d}}$ satisfy $\hat{Z}^{a\cdots b}{_{c\dots d}}= \delta^a{_\a} \cdots \delta^b{_\b}\, \delta^\c{_c}\cdots \delta^\d{_d} \,\hat{Z}^{\a\cdots \b}{_{\c\dots \d}}$, so there is no difference of using the Greek or  Latin indices for any tensor-field components at the original point $x'$. In the following, we will adapt the Greek indices on any tensor-field components at $x'$.

\section{Momentum-space representation of the Feynman propagator of a scalar field} \label{4}

From Eq. (\ref{seq-1}), one can show that the Feynamm Green's function of a scalar field satisfies \cite{Birrell-Davies,DeWitt-65}
\bea
&&\sqrt{|g(x)|} \,\left[- \star^{-1} d \star d   + m^2 + \xi R \right] G(x, x') = \,\delta (x-x'), \nn\\ 
&&\label{fp-1}
\eea where $|g(x)|\equiv |\det{g_{ab}}(x)|$. It is useful to define $\bar{G}(x, x')$ by
\bea
\bar{G}(x, x') = |g(x)|^{1/4}\, G(x, x')\, |g(x')|^{1/4}, 
\eea and Eq. (\ref{fp-1}) becomes
\bea
\left[\,(- |g(x)|^{1/4} \, \star^{-1} d \star d\, |g(x)|^{-1/4})  +  m^2 + \xi R \right] \, \bar{G}(x, x') 
= \,\delta (x-x'). \label{fp-2}
\eea It is known that, in the coincident limit $x \rightarrow x'$, the divergences of $G(x, x')$  and also effective action $W$ come from the high frequency field behavior \cite{Christensen-76, Birrell-Davies}. In the following, we will use $[\,Z\,]$ to denote the coincident limit of any two-point function $Z(x, x')$, i.e. $[\,Z\,] =\lim_{x\rightarrow x'} Z(x, x')$. These ultraviolet divergences can be obtained by solving Eq. (\ref{fp-2}) in the generalized Reimann-normal coordinates with asymptotic expansion in large wave number $k$.  

Eq. (\ref{fp-2}) in the generalized Riemann-normal coordinates $x^\alpha$ with associated orthonormal co-frame $\{e^a\}$ gives
\bea
&&(\eta^{\mu\nu} + \up{(1)}{\mathcal{F}}{^{\mu\nu}}{_{\alpha}} x^\alpha + \up{(2)}{\mathcal{F}}{^{\mu\nu}}{_{\alpha\beta}} x^\alpha x^\beta + \up{(3)}{\mathcal{F}}{^{\mu\nu}}{_{\alpha\beta\gamma}} x^\alpha x^\beta x^\gamma + \up{(4)}{\mathcal{F}}{^{\mu\nu}}{_{\alpha\beta\gamma\lambda}} x^\alpha x^\beta x^\gamma x^\lambda ) \partial_\mu \partial_\nu \bar{G}- m^2  \bar{G} \nn\\
&& + (\up{(1)}{\mathcal{S}}{^{\nu}}  + \up{(2)}{\mathcal{S}}{^{\nu}}{_{\alpha}} x^\alpha+ \up{(3)}{\mathcal{S}}{^{\nu}}{_{\alpha\beta}} x^\alpha x^\beta + \up{(4)}{\mathcal{S}}{^{\nu}}{_{\alpha\beta\gamma}} x^\alpha x^\beta x^\gamma ) \partial_\nu \bar{G} + [\, (\up{(2)}{\mathcal{P}} - \xi \hat{R}) \nn\\
&&+ ( \up{(3)}{\mathcal{P}}{_\alpha} - \xi \widehat{\nabla_\alpha R} )\, x^\alpha  + (\up{(4)}{\mathcal{P}}{_{\alpha\beta}}- \frac{\xi}{2} \widehat{\nabla_\beta \nabla_\alpha R} ) \,x^\alpha x^\beta\,] \,\bar{G}= -\delta(x), \label{scf-1}
\eea   \end{widetext}
 where the coefficients $\up{(i)}{\mathcal{F}}{^{\mu\nu}}{_{\cdots}}$, $\up{(i)}{\mathcal{S}}{^\nu}{_{\cdots}}$, and $\up{(i)}{\mathcal{P}}{_{\cdots}}$ involve the $i$ derivatives of orthonormal co-frame.
  The explicit expressions of these coefficients in terms of $\hat{T}^a{_{bc}}$, $\hat{R}^a{_{bcd}}$ and their covariant derivatives are given in Appendix. We have only retained the coefficients for $i\leqslant 4$ in Eq. (\ref{scf-1}) since the divergences of $W$ involve the coefficients up to four derivatives of $\{e^a\}$. However, the coefficients for $i =4$ become much complicated, the discussion of renormalization of $W$ will be restricted in \textit{totally anti-symmetric} torsion, i.e. $T_{abc} = T_{[abc]}$, where square brackets indicates index anti-symmetrization.  On the other hand, the divergences of $[\,G\,]$, which are used to study the renormalizability of interacting scalar fields, involve the coefficients for $i\leqslant 2$, so we \textit{do not} put any restriction on torsion for finding those divergences.     
 
 By making the $n$-dimensional Fourier transformation, $\bar{G}(x, x')$ in the momentum space yields
 \bea
 \bar{G}(x, x') =\int \frac{d^n k}{(2\pi)^n} e^{ik_\alpha x^\alpha} \bar{G}(k), \label{G-ft}
 \eea where $\bar{G}(k) = \bar{G}(k; x')$ is assumed to have compact support in the normal neighborhood of $x'$.  We consider the following expansion of $\bar{G}(k)$
 \bea
 \bar{G}(k)= \bar{G}_{0}(k) + \bar{G}_{1}(k) + \bar{G}_{2}(k) + \cdots, \label{G-expansion}
 \eea and 
 \bea
 \bar{G}_i (x, x') = \int \frac{d^n k}{(2\pi)^n} e^{ik_\alpha x^\alpha} \bar{G}_i (k), \label{G-expansion-1}
 \eea where $\bar{G}_{i}(k)$ involves the coefficients $\up{(i)}{\mathcal{F}}{^{\mu\nu}}{_{\cdots}}$, $\up{(i)}{\mathcal{S}}{^\nu}{_{\cdots}}$, and $\up{(i)}{\mathcal{P}}{_{\cdots}}$. For $i=0$, we have $\up{(0)}{\mathcal{F}}{^{\mu\nu}}{_{\cdots}}= \up{(0)}{\mathcal{S}}{^\nu}{_{\cdots}}= \up{(0)}{\mathcal{P}}{_{\cdots}}=0$. On dimensional ground, $G_i (k)$ is of order $k^{-(2 + i)}$ so Eq. (\ref{G-expansion}) corresponds to an asymptotic expansion of $\bar{G}(k)$ in large $k$ \cite{Bunch-Parker-79}. 
 
 To find the divergences of $[\,G\,]$, we first solve $\bar{G}_i$ for $i \leqslant 2$. By substituting Eq. (\ref{G-expansion-1}) into Eq. (\ref{scf-1}), the lowest-order equation yields
 \bea
 \eta^{\mu\nu}\partial_\mu \partial_\nu \bar{G}_0 - m^2 = -\delta (x), \label{G0-0}
 \eea which has the Minkowski-space solution
 \bea
 \bar{G}_0 (k) = \frac{1}{k^2 + m^2}. \label{G0}
 \eea From Eq. (\ref{G0-0}), we also know that $\bar{G}_0 (x, x')$ is a function of $\eta_{\mu\nu} x^\mu x^\nu \equiv x_\nu x^\nu$, i.e. Lorentz invariant. The equation for $\bar{G}_1(x, x')$ gives
 \bea
 \eta^{\mu\nu}\partial_\mu \partial_\nu \bar{G}_1 - m^2 \bar{G}_1 + \up{(1)}{\mathcal{F}}{^{\mu\nu}}{_\alpha} x^\alpha \partial_\mu \partial_\nu \bar{G}_0  + \up{(1)}{\mathcal{S}}{^{\nu}} \partial_\nu \bar{G}_0 =0. \nn\\ \label{G1-0}
 \eea Substituting the solution $\bar{G}_0(x, x')$ into Eq. (\ref{G1-0}) and using Eqs. (\ref{f1}), (\ref{s1}), we obtain
 \bea
 \bar{G}_1(k) = - \frac{i}{4}\, \hat{T}_\alpha \,\partial^\alpha \left(\frac{1}{k^2 + m^2}\right), \label{G1}
 \eea where $\hat{T}_\alpha = \hat{T}^{\beta}{_{\beta\alpha}}$ is the trace torsion, and $\partial^\alpha \equiv \partial/\partial k_\alpha$. It turns out that $\bar{G}_1(k)=0$ in the pseudo-Riemannian geometry, which has been shown in \cite{Bunch-Parker-79}. Using integrating by part, one can show that $[ \bar{G}_1 ] =0$ (see Sec. \ref{5}).
 
The equation for $\bar{G}_2(x, x')$ gives
%
\bea
&&
\eta^{\mu\nu}\partial_\mu \partial_\nu \bar{G}_2 - m^2 \bar{G}_2 + \up{(1)}{\mathcal{F}}{^{\mu\nu}}{_{\alpha}} x^\alpha \partial_\mu \partial_\nu \bar{G}_1 + \up{(1)}{\mathcal{S}}{^{\nu}} \partial_\nu \bar{G}_1 
\nn\\&&
+ \up{(2)}{\mathcal{F}}{^{\mu\nu}}{_{\alpha\beta}} x^\alpha x^\beta \partial_\mu \partial_\nu \bar{G}_0 + \up{(2)}{\mathcal{S}}{^{\nu}}{_{\alpha}} x^\alpha  \partial_\nu \bar{G}_0 
\nn\\&&
+ (\up{(2)}{\mathcal{P}} - \xi \hat{R}) \bar{G}_0 =0. \label{G2}
\eea By substituting the solutions $\bar{G}_0 (x, x')$, $\bar{G}_1 (x, x')$, Eqs. (\ref{f1})-(\ref{p2}) into Eq. (\ref{G2}) and integrating by part, a straightforward but tedious calculation yields
\begin{widetext}
\bea
\bar{G}_2 (k)&=&\left[(\frac{1}{6}-\xi) \hat{R} -\frac{1}{4} \hat{T}_\alpha \hat{T}^\alpha + \frac{1}{3} \widehat{\nabla_\alpha T^\alpha} -\frac{1}{8} \hat{T}_{\alpha\beta\gamma}\hat{T}^{\alpha\beta\gamma} -\frac{1}{6} \hat{T}_{\alpha\beta\gamma} \hat{T}^{\gamma\beta\alpha} \right] \,\frac{1}{(k^2+m^2)^2} \nn\\
&&- \frac{1}{8}\left[\frac{1}{4} \hat{T}_\alpha \hat{T}_\beta - \frac{1}{2}\hat{T}_{\alpha\beta\mu} \hat{T}^\mu + 4 \,( \up{(2)}{\mathcal{F}}{_\alpha}{^\mu}{_{(\beta\mu)}} + \up{(2)}{\mathcal{F}}{^\mu}{_\alpha}{_{(\beta\mu)}}\,) + 2 \up{(2)}{\mathcal{F}}{_{\alpha\beta}}{^{\mu}}{_{\mu}}  - 2 \up{(2)}{\mathcal{S}}{_{\alpha\beta}} \right] \partial^\alpha \partial^\beta \left( \frac{1}{k^2 + m^2}\right), \label{G2-1}
\eea 
 \end{widetext}   where the indices are up and lower by $\eta^{\mu\nu}$ and $\eta_{\mu\nu}$ and round brackets indicate index symmetrization. In Sec. \ref{5}, it will be shown that the second line of Eq. (\ref{G2-1}) does not contribute to $[\,\bar{G}_2\,]$. In pseudo-Riemannian geometry, Eq. (\ref{G2-1}) reduces to
 \bea
 \bar{G}_2 (k)=   \frac{(\frac{1}{6}-\xi)\hat{\tilde{R}}}{(k^2+m^2)^2}, 
 \eea which is the same as in \cite{Bunch-Parker-79}.


\subsection{A special case: the total antisymmetric torsion} \label{4-1}

In this subsection, we will consider background torsion to be totally anti-symmetric and find the divergences of effective action $W$ in this restricted background gravitaional fields. The reason is that the totally anti-symmetric torsion plays a significant role for generating inflation in the early Universe \cite{Wang-Wu-09}. Moreover, since it is necessary to obtain $\bar{G}_4 (k)$ for finding the divergences of $W$, this consideration will largely simplify our calculations. 

When $T_{\alpha\beta\gamma} = T_{[\alpha\beta\gamma]}$, the autoparallels and geodesics will coincide, and it gives $\up{(1)}{\mathcal{F}}{^{\mu\nu}}{_{\alpha}}= \up{(1)}{\mathcal{S}}{^{\nu}}=0$. So Eq. (\ref{G1-0}) gives a trivial solution $\bar{G}_1 (k) = 0$. Since $\bar{G}_0 (x, x')$ is a function of $x_\mu x^\nu$, and using Eqs. (\ref{f2-1})-(\ref{s2-1}), we obtain 
\bea
\up{(2)}{\mathcal{F}}{^{\mu\nu}}{_{\alpha\beta}} x^\alpha x^\beta \partial_\mu \partial_\nu \bar{G}_0 + \up{(2)}{\mathcal{S}}{^{\nu}}{_{\alpha}} x^\alpha  \partial_\nu \bar{G}_0 \equiv 0. \label{G0-id}
\eea Therefore, Eq. (\ref{G2}) becomes
\bea
\eta^{\mu\nu}\partial_\mu \partial_\nu \bar{G}_2 - m^2 \bar{G}_2 + (\up{(2)}{\mathcal{P}} - \xi \hat{R}) \bar{G}_0 =0, \label{G2-2}
\eea which has a solution
\bea
\bar{G}_2 (k) = \left[(\frac{1}{6}-\xi) \hat{R} + \frac{1}{24} \hat{T}_{\alpha\beta\gamma}\hat{T}^{\alpha\beta\gamma}  \right] \,\frac{1}{(k^2+m^2)^2}. \label{G2_2}
\eea 
Eq. (\ref{G2-2}) indicates that $\bar{G}_2 (x, x')$ is Lorentz invariant and hence it is also a function of $x^\alpha x_\alpha$. It follows that $\bar{G}_2 (x, x')$ also satisfies Eq. (\ref{G0-id}). Moreover, by using Eqs. (\ref{f3}), (\ref{s3}), (\ref{f4}), (\ref{s4}), a straightforward but tedious calculation gives two more identities
\bea
&&\up{(3)}{\mathcal{F}}{^{\mu\nu}}{_{\alpha\beta\gamma}} x^\alpha x^\beta x^\gamma \partial_\mu \partial_\nu \bar{G}_0 + \up{(3)}{\mathcal{S}}{^{\nu}}{_{\alpha\beta}} x^\alpha x^\beta \partial_\nu \bar{G}_0 \equiv 0, \\
&&\up{(4)}{\mathcal{F}}{^{\mu\nu}}{_{\alpha\beta\gamma\lambda}} x^\alpha x^\beta x^\gamma x^\lambda \partial_\mu \partial_\nu \bar{G}_0 + \up{(4)}{\mathcal{S}}{^{\nu}}{_{\alpha\beta\gamma}} x^\alpha x^\beta x^\gamma \partial_\nu \bar{G}_0 \equiv 0. \nn\\
\eea 
so $\bar{G}_3 (x, x')$ and $\bar{G}_4 (x, x')$ satisfy
\begin{widetext}
\bea
&&\eta^{\mu\nu}\partial_\mu \partial_\nu \bar{G}_3 - m^2 \bar{G}_3 + ( \up{(3)}{\mathcal{P}}{_\alpha} - \xi \widehat{\nabla_\alpha R} )\, x^\alpha  \bar{G}_0= 0. \label{G3}\\
&& \eta^{\mu\nu}\partial_\mu \partial_\nu \bar{G}_4 - m^2 \bar{G}_4 +(\up{(2)}{\mathcal{P}} -\, \xi \hat{R}) \,\bar{G}_2+ ( \up{(4)}{\mathcal{P}}{_{\alpha\beta}} - \frac{\xi}{2} \widehat{\nabla_\beta \nabla_\alpha R} )\, x^\alpha x^\beta \bar{G}_0= 0. \label{G4}
\eea By substituting Eq. (\ref{p3}) into Eq. (\ref{G3}) and integrating by part, we obtain
\bea
\bar{G}_3(k) = \frac{i}{2}\left[(\frac{1}{12} -\xi) \widehat{\nabla_\alpha R} +  \frac{1}{12}(2 \widehat{\nabla^\beta{R}{_{(\beta\alpha)}}}+2 \hat{R}^\gamma{_{(\alpha\beta)\lambda}}\hat{T}^{\lambda\beta}{_{\gamma}} 
 + \hat{T}^{\beta\lambda\gamma} \nabla_{(\alpha} \hat{T}_{\lambda)}{_{\gamma\beta}}+ \frac{1}{2}\hat{T}_{\alpha\gamma\beta} \nabla_\lambda \hat{T}^{\lambda\gamma\beta} )\right ] \partial^\alpha \frac{1}{(k^2 + m^2)^2}, \label{G3-1}
\eea 
where $\nabla^\alpha \equiv g^{\alpha\beta} \nabla_\beta$. When torsion vanishes, Eq. (\ref{G3-1}) reduces to 
\bea
\bar{G}_3(k) = \frac{i}{2} (\frac{1}{6} -\xi) \widehat{\tilde{\nabla}_\alpha \tilde{R}} \,\,\partial^\alpha \frac{1}{(k^2 + m^2)^2}, 
\eea where the Bianchi identities have been used. It agrees with the result in \cite{Bunch-Parker-79}. Similarly, substituting Eqs. (\ref{p2-1}) and (\ref{p4}) into Eq. (\ref{G4}) and integrating by part yields
\bea
\bar{G}_4 (k)& =&  \left[(\frac{1}{6}-\xi) \hat{R} + \frac{1}{24} \hat{T}_{\alpha\beta\gamma}\hat{T}^{\alpha\beta\gamma}  \right]^2 \frac{1}{(k^2+m^2)^3}+ \frac{2}{3} \left(\up{(4)}{\mathcal{P}}{^\alpha}{_\alpha} - \frac{1}{2}\xi\, \widehat{\square R} \right) \frac{1}{(k^2+m^2)^3} \nn\\
&& - \frac{1}{3}\left( \up{(4)}{\mathcal{P}}{_{\alpha\beta}} - \frac{\xi}{2} \widehat{\nabla_\beta \nabla_\alpha R} \right) \partial^\alpha \partial^\beta\,  \frac{1}{(k^2+m^2)^2}, \label{G4-1}
\eea where $\square \equiv \nabla^\alpha \nabla_\alpha$ and
\bea
\up{(4)}{\mathcal{P}}{^\alpha}{_\alpha} &=& \frac{1}{20} \widehat{\square R}+ \frac{1}{10} \widehat{\nabla_{(\beta}\nabla_{\alpha)}R^{\alpha\beta}}-\frac{1}{72}\hat{R}_{\alpha\beta}\hat{R}^{\alpha\beta} -\frac{1}{360}\hat{R}_{\alpha\beta}\hat{R}^{\beta\alpha}
-\frac{1}{90}\hat{R}^\gamma{_\lambda} \widehat{\nabla_\alpha T^{\lambda\alpha}{_\gamma}} -\frac{1}{30}\hat{R}^\gamma{_\lambda} \hat{T}^{\lambda\alpha}{_\mu} \hat{T}^\mu{_{\alpha\gamma}} \nn\\
&&-\frac{3}{10}  \widehat{\nabla_{(\nu}\nabla_{\alpha)}T^{\gamma\alpha}{_\lambda}} \,\, \hat{T}^{\lambda\nu}{_\gamma}
-\frac{1}{45}\widehat{\nabla_\beta T^{\gamma\beta}{_\lambda}}\, \widehat{\nabla_\nu T^{\lambda\nu}{_\gamma}}+ \frac{1}{45}\widehat{\nabla_\nu T^{\gamma\beta\lambda}} \,\widehat{\nabla^\nu T_{\gamma\beta\lambda}} -\frac{1}{45}\widehat{\nabla^\nu T^{\gamma\beta}{_\lambda}}\, \widehat{\nabla_\beta T^{\lambda}{_{\nu\gamma}}}\nn\\
&&+\frac{11}{180}\hat{R}_{\kappa\nu\beta\gamma}\hat{T}^{\kappa\nu\lambda} \hat{T}^{\beta\gamma}{_{\lambda}}
-\frac{11}{180}\hat{R}^{\lambda\nu\kappa\beta}\hat{T}_{\lambda\beta\gamma} \hat{T}_{\nu\kappa}{^{\gamma}}
-\frac{1}{180}\hat{R}^{\gamma}{_\beta}{^\nu}{_\lambda} \widehat{\nabla_{(\nu}T^{\lambda\beta}{_{\gamma)}}} +\frac{1}{90}\hat{R}^{\gamma\beta\lambda\alpha}\hat{R}_{\gamma\lambda\beta\alpha} +\frac{1}{90}\hat{R}^{\gamma\beta\lambda\alpha}\hat{R}_{\lambda\alpha\gamma\beta}\nn\\
&&-\frac{1}{2880}\hat{T}^{\lambda\beta\gamma}\hat{T}_{\lambda\beta\kappa}\hat{T}_{\alpha\nu}{^\kappa} \hat{T}^{\alpha\nu}{_\gamma} +\frac{1}{1440}\hat{T}^{\gamma\beta}{_\nu} \hat{T}^{\nu\lambda\alpha}\hat{T}_{\alpha\beta\kappa} \hat{T}^\kappa{_{\lambda\gamma}}. \label{G4-2}
\eea Sec. \ref{5} will show that $[\,\bar{G}_3\,]=0$ and the second line of Eq. (\ref{G4-1}) does not contribute to $[\,\bar{G}_4\,]$. When torsion vanishes, Eq. (\ref{G4-1}) becomes
\bea
\bar{G}_4 (k) &=&  \left[(\frac{1}{6}-\xi)^2 \hat{\tilde{R}}^2 +\frac{1}{3}(\frac{1}{5}- \xi ) \widehat{\tilde{\square} \tilde{R}} - \frac{1}{90} \hat{\tilde{R}}_{\alpha\beta} \hat{\tilde{R}}^{\alpha\beta}+ \frac{1}{90}\hat{\tilde{R}}^{\gamma\beta\lambda\alpha}\,\hat{\tilde{R}}_{\gamma\beta\lambda\alpha} \right] \frac{1}{(k^2+m^2)^3} \\
 &+& \left[ \frac{1}{6}(\xi - \frac{3}{20} ) \widehat{\tilde{\nabla}_\alpha \tilde{\nabla}_\beta \tilde{R}} - \frac{1}{120} \widehat{\tilde{\square} \tilde{R}_{\alpha\beta}} +  \frac{1}{90} \hat{\tilde{R}}_{\alpha\lambda} \hat{\tilde{R}}^{\lambda}{_\beta}+ \frac{1}{270} \hat{\tilde{R}}_{\gamma\lambda} \hat{\tilde{R}}{_\alpha}{^{\gamma\lambda}}{_\beta} - \frac{1}{180}\hat{\tilde{R}}^{\gamma\kappa\lambda\alpha}\,\hat{\tilde{R}}_{\gamma\kappa\lambda\beta} \right] \partial^\alpha \partial^\beta\,  \frac{1}{(k^2+m^2)^2}, \nn
\eea 
which agrees with the result in \cite{Bunch-Parker-79}. The Feynamm propagator can be obtained by giving $m^2$ an infinitesimal negative imaginary part $i\epsilon$, i.e. $m^2 + i\epsilon$, and take $i \epsilon$ to be zero at the end of calculation. Since our calculations are valid in $n$ dimensions, it is natural to use dimensional regularization to handle the divergences of Feynamm propagator and effective action in the coincident limit. 



\section{Renormalization of a scalar field in Riemann-Cartan spacetime} \label{5}

It is known that proper-time representation can be derived from momentum-space representation in the $n$ dimensional pseudo-Riemannian structure of space-time \cite{Bunch-Parker-79}. We will show that the derivation can be extended to $n$ dimensional Riemann-Cartan spacetime. In the following, we only consider the approximate solution of $G(x, x')$ up to $G_4 (x, x')$. Substituting Eqs. (\ref{G0}), (\ref{G1}), (\ref{G2-1}), (\ref{G3-1}), (\ref{G4-1}) into (\ref{G-ft}) and integrating by part yields 
%
\bea
 \bar{G}(x, x') =\int \frac{d^n k}{(2\pi)^n} e^{ik_\alpha x^\alpha} \left[ 1 - \frac{1}{4} \hat{T}_\alpha x^\alpha + {a}_{{\alpha\beta}} x^\alpha x^\beta + ( a+ b_\alpha x^\alpha + c_{\alpha\beta} x^\alpha x^\beta ) (-\frac{\partial}{\partial m^2}) +  c\,\, (\frac{\partial}{\partial m^2})^2\right] \frac{1}{k^2 + m^2}, \label{G(x)}
\eea where
\bea
a_{\alpha\beta}&=& \frac{1}{8}\left[\frac{1}{4} \hat{T}_\alpha \hat{T}_\beta - \frac{1}{2}\hat{T}_{\alpha\beta\mu} \hat{T}^\mu + 4 \,( \up{(2)}{\mathcal{F}}{_\alpha}{^\mu}{_{(\beta\mu)}} + \up{(2)}{\mathcal{F}}{^\mu}{_\alpha}{_{(\beta\mu)}}\,) + 2 \up{(2)}{\mathcal{F}}{_{\alpha\beta}}{^{\mu}}{_{\mu}}  - 2 \up{(2)}{\mathcal{S}}{_{\alpha\beta}} \right],\\
a &=&  (\frac{1}{6}-\xi) \hat{R} -\frac{1}{4} \hat{T}_\alpha \hat{T}^\alpha + \frac{1}{3} \widehat{\nabla_\alpha T^\alpha} -\frac{1}{8} \hat{T}_{\alpha\beta\gamma}\hat{T}^{\alpha\beta\gamma} -\frac{1}{6} \hat{T}_{\alpha\beta\gamma} \hat{T}^{\gamma\beta\alpha},\\
b_\alpha &=&  \frac{i}{2}\left[(\frac{1}{12} -\xi) \widehat{\nabla_\alpha R} +  \frac{1}{12}(2 \widehat{\nabla^\beta{R}{_{(\beta\alpha)}}}+2 \hat{R}{_{\gamma (\alpha\beta)\lambda}}\hat{T}^{[\lambda\beta\gamma]} 
 + \hat{T}^{[\beta\lambda\gamma]} \nabla_{(\alpha} \hat{T}_{[\lambda)}{_{\gamma\beta]}}+ \frac{1}{2}\hat{T}_{[\alpha\gamma\beta]} \nabla_\lambda \hat{T}^{[\lambda\gamma\beta]} )\right ], \\
c_{\alpha\beta}&=&  \frac{1}{3}\left( \up{(4)}{\mathcal{P}}{_{\alpha\beta}} - \frac{\xi}{2} \widehat{\nabla_\beta \nabla_\alpha R} \right), \\
c &=& \frac{1}{2} \left[(\frac{1}{6}-\xi) \hat{R} + \frac{1}{24} \hat{T}_{[\alpha\beta\gamma]}\hat{T}^{[\alpha\beta\gamma]}\right]^2 + \frac{1}{3} \left(\up{(4)}{\mathcal{P}}{^\alpha}{_\alpha} - \frac{1}{2}\xi\, \widehat{\square R} \right). 
\eea 
%
\end{widetext}  
It should be pointed out that the coefficients $a$ and $a_{\alpha\beta}$ are considered in a general background torsion field but the other coefficients $b_\alpha$, $c_{\alpha\beta}$ and $c$ are considered in a background totally anti-symmetric torsion field.  

Defining 
\bea
F(x, x'; is)& =&  1 - \frac{1}{4} \hat{T}_\alpha x^\alpha + {a}_{{\alpha\beta}} x^\alpha x^\beta \nn\\
   &+& ( a+ b_\alpha x^\alpha + c_{\alpha\beta} x^\alpha x^\beta )i s + c (i s)^2, 
\eea and
using the integral representation
\bea
( k^2 + m^2 + i \epsilon )^{-1} = \int_0^\infty i ds \exp [- is (k^2 + m^2 + i \epsilon)],  
\eea one then perform $d^n k$ integration in Eq. (\ref{G(x)}) to obtain (dropping $i\epsilon$)
\bea
\bar{G}(x, x') &=& i (4\pi)^{-n/2} \int_0^\infty i ds (i s)^{-n/2} \nn\\
&&\times \exp [- i m^2 s - (\sigma/2 i s)] F(x, x'; is),
\eea where $\sigma(x, x') = \frac{1}{2} x^\alpha x_\alpha$ is half square of the autoparallel distance between $x$ and $x'$. Since $|g(x')|=1$ in the generalized Riemann-normal coordinates, it gives 
\bea
G(x, x') = |g(x)|^{-1/4}\bar{G} (x, x'). 
\eea By introducing a determinant defined by\footnote{Eq. (\ref{VVD}) returns to the well-known Van Vleck determinant in pseudo-Riemannian geometry.}  
\bea
\bigtriangleup (x, x') = - |g(x)|^{-1/2
} \det [-\partial_\mu \partial_{\nu '} \sigma] |g(x')|^{-1/2} \label{VVD}
\eea and noticing that Eq. (\ref{VVD}) reduces to $|g(x)|^{-1/2}$ in the generalized Riemann-normal coordinates, we obtain 
\bea
G (x, x') &=&  \frac{i\bigtriangleup^{1/2}(x, x')}{(4\pi)^{n/2}}  \int_0^\infty i ds (i s)^{-n/2} \nn\\
&&\times \exp [- i m^2 s - (\sigma/2 i s)] F(x, x'; is), \label{G}
\eea which may be considered as the proper-time representation in $n$ dimensional Riemann-Cartan spacetime. When torsion vanishes, Eq. (\ref{G}) yields the usual expression of DeWitt-Schwinger proper-time representation in $n$ dimensional pseudo-Riemannian structure of space-time.

It is known that the first $\frac{n}{2} $ terms of Eq. (\ref{G}) are divergent at $x \rightarrow x'$ limit \cite{Birrell-Davies}. If one considers that $n$ can be analytically continued throughout the complex plane, Eq. (\ref{G}) at $x \rightarrow x'$ limit becomes
\bea
G (x, x) &=& \frac{i}{(4\pi)^{n/2}}\Big[ m^2 \Gamma(-\frac{n}{2}+1) + a(x) \Gamma(-\frac{n}{2} + 2) \nn \\
&&+ m^{-2} c(x) \Gamma(-\frac{n}{2} + 3)\Big]. \label{G(x,x)}
\eea When $n \rightarrow 4$, Eq. (\ref{G(x,x)}) indicates that only the first two terms are divergent. 

From Eq. (\ref{EA}), it can be shown that \cite{Birrell-Davies}
\bea
W = - \frac{i}{2}\int_M \left[\lim_{x\rightarrow x'} \int_0^\infty i ds (is)^{-1} G(x, x')\right] \star 1. \label{L-eff} 
\eea By substituting Eq. (\ref{G}) into Eq. (\ref{L-eff}), the divergent terms in the four dimensional spacetime yield
\bea
L_{div} &=&\lim_{n\rightarrow 4} \frac{1}{(32 \pi^2)}\Big[ m^4 \Gamma(-\frac{n}{2}) + m^2 a(x) \Gamma(-\frac{n}{2} + 1) \nn \\
&&+  c(x) \Gamma(-\frac{n}{2} + 2)\Big].
\eea It turns out that the divergent terms are entirely geometrical and involve only $a(x)$ and $c(x)$.  By adding the counterterms, which contain bare coefficients,  into the gravitational Lagrangian, the infinite quantities of $L_{div}$ can be absorbed into bare coefficients to obtain renormalized physical quantities. It should be pointed out that, for totally anti-symmetric torsion, $a(x)$ and $c(x)$ may be compared to the coefficients $b_2$ and $b_4$ (i.e. Eq. (4.2.27) and and (4.3.10)) in \cite{Goldthorpe-80}. It is easy to see that $a(x)$ in totally anti-symmetric torsion case, which is referred to Eq. (\ref{G2_2}), is equivalent to $b_2$. However, we have not verified the equivalence of $c(x)$ and $b_4$ yet, since it involves using the Bianchi identities.


\section{Conclusion and Discussion}

  We obtain the momentum-space representation of the Feynamm propagator of a free massive scalar field in Riemann-Cartan spacetime. Moreover, the proper-time representation in $n$ dimensional Riemann-Cartan spacetime has been derived from our momentum-space representation. It leads us to find the divergences of the one-loop effective action by using dimensional regularization. It turns out that the divergences of one-loop effective action of the scalar field are purely geometrical and involve full curvature, torsion and their covariant derivative. It is interesting to notice that though there is no direct coupling between torsion and the scalar field in the classical action, those divergences do contain torsion parts. When torsion vanishes, our momentum-space representation agrees with the results in \cite{Bunch-Parker-79}. 

It has been demonstrated that momentum-space representation is useful for studying the renormalizability of interacting fields in pseudo-Riemannian structure of spacetime \cite{Bunch-Parker-79}. So our current work may also be useful for studying renormalizability of interacting scalar fields in Riemann-Cartan spacetime. Moreover, finding momentum-space representation of Feynamm propagator for spin 1/2 field in Riemann-Cartan spacetime is straightforward by using the generalized Riemann-normal coordinates. These considerations will be our future work.

Our original motivation is to study quantum effects of our inflation model  \cite{Wang-Wu-09} in Riemann-Cartan spacetime. It turns out that our inflation model, which contain quadratic curvature terms, is a subclass of the effective action.  Therefore, it might be interesting to find the renormalized stress 3-forms and spin 3-forms, and study these quantum effects in the early Universe. A further investigation on reheating and primordial perturbations  will also be studied in the future.

\acknowledgments

CHW would like to thank Prof Hing-Tong Cho and Prof Chopin Soo for helpful discussions and comments. YHW was suppoerted by Center for Mathematics and Theoretical Physics, National Central University. CHW was supported by the National Science Council of the Republic of China under the grants NSC 96-2112-M-032-006-MY3 and 98-2811-M-032-014. 

\begin{widetext}
\begin{appendix}

\section{Feynman propagator in the generalized Riemann-normal coordinates} \label{A}

In Sec. \ref{3}, we obtained the orthonormal co-frames $\{e^a\}$ and connection 1-forms $\{\omega^a{_b}\}$ in the generalized Riemann-normal coordinates. To obtain Eq. (\ref{fp-2}) in the generalized Riemann-normal coordinates, it is useful to find the metric components $g_{\alpha\beta}$ with respect to $\{d x^\alpha\}$. Using
\bea
g= \eta_{ab}\, e^a \otimes e^b = g_{\alpha\beta}\, d x^\alpha \otimes d x^\beta \label{g}
\eea and substituting Eqs. (\ref{e1}), (\ref{e2}), (\ref{e3}) into (\ref{g}) gives
\bea
g_{\alpha\beta} = \eta_{\alpha\beta} -  \hat{T}_{(\a\b)\c}x^\c + \frac{1}{3} \left[\hat{R}_{\c(\a\b)\d} -2\widehat{\nabla_\d T}_{(\a\b)\c}
  +\half (\hat{T}_{\a\c\e} \hat{T}^\e\,_{\d\b}+\hat{T}_{\b\c\e}\hat{T}^\e\,_{\d\a}) +\frac{3}{4} \hat{T}^\e\,_{\c\a}\hat{T}_{\e\d\b} \right]  x^\c x^\d +  \cdots, \label{g-1}
\eea which can be used to find the solutions $G_0(x, x')$, $G_1(x, x')$ and $G_2(x, x')$. However, the solutions $G_3(x,x')$ and $G_4(x, x')$ are restricted in the background totally anti-symmetric torsion $T_{\a\b\c} = T_{[\a\b\c]}$, so substituting Eqs. (\ref{e1}), (\ref{e2}), (\ref{e3}), (\ref{e4}), (\ref{e5}) into (\ref{g}) and considering torsion field to be totally anti-symmetric yields
\bea
g_{\a\b} &=& \eta_{\a\b} +  \frac{1}{3} \left[\hat{R}_{\c(\a\b)\d} + \frac{1}{4} \hat{T}_{\a\c\e} \hat{T}^\e{_{\d\b}}\right] x^\c x^\d + \frac{1}{12}\Big[-\widehat{\nabla_\d R}_{\a\c\b\e} +\half \hat{R}_{\a\c\e}{^\f} \hat{T}_{\f\d\b} 
+\half \hat{T}_{\a\c\f} \widehat{\nabla_\d T}{^\f}{_{\e\b}}\nn\\
 &-& \half \hat{T}_{\a\c\f} \hat{R}^\f\,_{\d\e\b} + \a \leftrightarrow \b \Big] x^\c x^\d x^\e +  \Big[\,\frac{1}{120}\big(-3 \widehat{\nabla_\f \nabla_\e R}_{\a\c\b\d} +3 \widehat{\nabla_\e R}_{\a\c\d\g} \hat{T}^\g{_{\f\b}} + \hat{R}_{\a\c\d\g} \hat{R}^\g{_{\e\f\b}}\nn\\
  &+& 2 \hat{R}_{\a\c\d\g} \widehat{\nabla_\e T}{^\g}{_{\f\b}}
  + \hat{R}_{\a\c\d\g} \hat{T}^\g\,_{\f\h} \hat{T}^\h\,_{\e\b} + 9 \widehat{\nabla_\e\nabla_\d T}_{\a\c\g} \hat{T}^\g\,_{\f\b}\big) 
  -\frac{1}{45} \widehat{\nabla_\d T}_{\a\c\g} \hat{R}^\g{_{\e\f\b}}
  + \frac{1}{90} \widehat{\nabla_\d T}_{\a\c\g} \widehat{\nabla_\f T}{^\g}{_{\e\b}} \nn\\
  &+& \frac{1}{360}\widehat{\nabla_\d T}_{\a\c\g} \hat{T}^\g\,_{\e\h} \hat{T}^\h{_{\f\b}}
  -\frac{1}{40} \hat{T}_{\a\c\g} \widehat{\nabla_\f R}{^\g}{_{\d\e\b}}
  - \frac{1}{80} \hat{T}_{\a\c\g} \hat{R}^\g{_{\e\f\h}} \hat{T}^\h{_{\d\b}}
   - \frac{1}{80} \hat{T}_{\a\c\g}\hat{T}^\g{_{\e\h}} \hat{R}^\h{_{\f\d\b}} \nn\\
   &+&\frac{1}{720} \hat{T}_{\a\c\g} \hat{T}^\g{_{\e\h}} \hat{T}^\h{_{\f\i}}\hat{T}^\i{_{\d\b}}
  +\frac{1}{72} \hat{R}_{\g\c\d\a} \hat{R}^\g{_{\e\f\b}}
  + \a \leftrightarrow \b \Big] x^\c x^\d x^\e x^\f + \cdots, \label{g_ab}
\eea where $\a \leftrightarrow \b$ denotes interchange of the indices. It is not difficult to verify that when torsion field vanishes, Eq. (\ref{g_ab}) will return to the well-known result obtained in the pseudo-Riemannian geometry \cite{Petrov-69}.  

Since Eq. (\ref{fp-2}) only involve exterior derivative $d$ acting on $\bar{G}$, it can be expressed in terms of $g_{\a\b}$ and Christoffel symbol $\tilde{\Gamma}{^{\a}}{_{\b\c}}$  
\bea g^{\a\b}\partial_\a \partial_\b \bar G + \partial_\a g^{\a\b} \partial_\b \bar G -
\left(\half \partial_\a g^{\a\b} \tilde{\Gamma}_{\c\b}{^\c} +\frac{1}{4} g^{\a\b} \tilde{\Gamma}_{\c\a}{^\c}\tilde{\Gamma}_{\d\b}{^\d} 
+ \half \tilde{\Gamma}_{\c\b}{^\c}{_{,\a}} g^{\a\b}\right) \bar G -(m+\xi R) \bar G = -\d (x-x'). \label{fp-3}
\eea In the generalized Riemann-normal coordinates, one has the following expansions
\bea
&&g^{\h\i}= \eta^{\mu\nu} + \up{(1)}{\mathcal{F}}{^{\mu\nu}}{_{\alpha}} x^\alpha + \up{(2)}{\mathcal{F}}{^{\mu\nu}}{_{\alpha\beta}} x^\alpha x^\beta + \up{(3)}{\mathcal{F}}{^{\mu\nu}}{_{\alpha\beta\gamma}} x^\alpha x^\beta x^\gamma +
\up{(4)}{\mathcal{F}}{^{\mu\nu}}{_{\alpha\beta\gamma\lambda}} x^\alpha x^\beta x^\gamma x^\lambda + \cdots \label{g-e}\\
&&\partial_\h g^{\h\i} = \up{(1)}{\mathcal{S}}{^{\nu}}  + \up{(2)}{\mathcal{S}}{^{\nu}}{_{\alpha}} x^\alpha+ \up{(3)}{\mathcal{S}}{^{\nu}}{_{\alpha\beta}} x^\alpha x^\beta + \up{(4)}{\mathcal{S}}{^{\nu}}{_{\alpha\beta\gamma}} x^\alpha x^\beta x^\gamma +\cdots \label{g-e1}\\
&&- \left(\half \partial_\a g^{\a\b} \tilde{\Gamma}_{\c\b}{^\c} +\frac{1}{4} g^{\a\b} \tilde{\Gamma}_{\c\a}{^\c}\tilde{\Gamma}_{\d\b}{^\d} 
+ \half \tilde{\Gamma}_{\c\b}{^\c}{_{,\a}} g^{\a\b}\right) =  \up{(2)}{\mathcal{P}}  +  \up{(3)}{\mathcal{P}}{_\alpha} \, x^\alpha  + \up{(4)}{\mathcal{P}}{_{\alpha\beta}}  \,x^\alpha x^\beta\,+ \cdots. \label{g-e2}
 \eea By substituting Eqs. (\ref{g-e})-(\ref{g-e2}) into Eq. (\ref{fp-3}), we then obtain Eq. (\ref{scf-1}).

Using Eq. (\ref{g-1}), we obtain
\bea
\up{(1)}{\mathcal{F}}{^{\h\i}}{_{\a}} &=& \hat{T}^{(\h\i)}{_\a}, \label{f1}\\
\up{(1)}{\mathcal{S}}{^{\i}} &=& \frac{1}{2}\,\hat{T}^{\i}, \label{s1}\\
\up{(2)}{\mathcal{F}}{^{\h\i}}{_{\a\b}}&=& \hat{T}_{\e}{^\h}{_\a} \hat{T}^{\e\i}{_\b} -\frac{1}{3} \left[\hat{R}_{\a}{^{(\h\i)}}{_\b} -2\widehat{\nabla_\b T}{^{(\h\i)}}{_\a}
  +\half \big(\hat{T}^{\h}{_{\a\e}} \hat{T}^\e{_\b}{^\i}+\hat{T}^{\i}{_{\a\e}}\hat{T}^\e{_\b}{^\h}\big) +\frac{3}{4} \hat{T}^\e{_{\a}}{^\h}\hat{T}_{\e\b}{^\i} \right], \label{f2}\\
\up{(2)}{\mathcal{S}}{^{\i}}{_{\a}}& = & \frac{2}{3}\,\hat{T}_{\e} \,\hat{T}^{\e\i}{_\a} + \frac{3}{4}\hat{T}_{\e}{^\h}{_\a} \hat{T}^{\e\i}{_\h} 
-\frac{1}{6} \big(\hat{R}_{\a}{^\i}+\hat{R}^{\i}{_{\a}}  - 4\widehat{\nabla_\h T}{^{(\h\i)}}{_\a}- 2 \widehat{\nabla_\a T}{^{\i}}
 +\hat{T}^{\i}{_{\h\e}}\hat{T}^\e{_\a}{^\h} + \hat{T}^{\h}{_{\a\e}}\hat{T}^\e{_\h}{^\i} \big), \label{s2}\\
\up{(2)}{\mathcal{P}}&=&-\frac{3}{16}\, \hat{T}^{\a}\,\hat{T}_{\a} + \frac{1}{6}\, \hat{R} + \frac{1}{3}\widehat{\nabla_\a T}{^{\a}}- \frac{1}{24}\hat{T}_{\a\b\c}\,\hat{T}^{\c\b\a}. \label{p2}
\eea
In the case of totally anti-symmetric torsion, one may use Eq. (\ref{g_ab}) to obtain
\bea
\up{(1)}{\mathcal{F}}{^{\h\i}}{_{\a}}&=& \up{(1)}{\mathcal{S}}{^{\i}} =0,\\
\up{(2)}{\mathcal{F}}{^{\h\i}}{_{\a\b}}&=& -  \frac{1}{3} \big(\hat{R}_{\a}{^{(\h\i)}}{_\b} + \frac{1}{4} \hat{T}{^\h}{_{\a\e}} \hat{T}^\e{_\b}{^\i}\big),\label{f2-1}\\
\up{(2)}{\mathcal{S}}{^{\i}}{_{\a}} &=&-\frac{1}{6} \big(\hat{R}_{\a}{^\i}+\hat{R}^{\i}{_\a}+ \frac{1}{2} \hat{T}_{\e\h\a}\hat{T}^{\e\h\i} \big),\label{s2-1}\\
\up{(2)}{\mathcal{P}}&=&  \frac{1}{6} \big(\hat{R} + \frac{1}{4} \hat{T}_{\e\h\a}\hat{T}^{\e\h\a} \big), \label{p2-1}\\
\up{(3)}{\mathcal{F}}{^{\h\i}}{_{\a\b\c}}&=& - \frac{1}{12}\Big(-\widehat{\nabla_\b R}{^\h}{_\a}{^\i}{_\c} +\half \hat{R}{^\h}{_{\a\c}}{^\f} \hat{T}_{\f\b}{^\i} 
+\half \hat{T}{^\h}{_{\a\f}} \widehat{\nabla_\b T}{^\f}{_\c}{^\i}
 - \half \hat{T}{^\h}{_{\a\f}} \hat{R}^\f{_{\b\c}}{^\i} + \h \leftrightarrow \i \Big),\label{f3}\\
 \up{(3)}{\mathcal{S}}{^{\i}}{_{\a\b}} &=& \up{(3)}{\mathcal{F}}{^{\i\h}}{_{\h\a\b}} + \up{(3)}{\mathcal{F}}{^{\i\h}}{_{\a\h\b}} + \up{(3)}{\mathcal{F}}{^{\i\h}}{_{\a\b\h}}, \label{s3}\\
 \up{(3)}{\mathcal{P}}{_\a} &=& \frac{1}{12}\Big(\widehat{\nabla_\a R}  + \widehat{\nabla_\h R}{^\h}{_\a} +\widehat{\nabla_\h R}{_\a}{^\h}+2\,\hat{R}{_{\h(\i\a)\f}} \hat{T}^{\f\i\h} 
-  \hat{T}{^{\h\i}}{_\f} \widehat{\nabla_{(\a} T}{^\f}{_{\i)\h}} - \half\, \hat{T}{^{\h\a\f}} \widehat{\nabla_\i T}{_{\f}}{^\i}{_\h}
   \Big), \label{p3}
   \eea
   \bea
  \up{(4)}{\mathcal{F}}{^{\h\i}}{_{\a\b\c\d}} &=& - \Big[\,\frac{1}{120}\big(-3 \widehat{\nabla_\d \nabla_\c R}{^\h}{_\a}{^\i}{_\b} +3 \widehat{\nabla_\c R}{^\h}{_{\a\b\g} \hat{T}^\g{_{\d}}{^\i}} + \hat{R}{^\h}{_{\a\b\g}} \hat{R}^\g{_{\c\d}}{^\i}
  + 2 \hat{R}{^\h}{_{\a\b\g}} \widehat{\nabla_\c T}{^\g}{_\d}{^\i}
  + \hat{R}{^\h}{_{\a\b\g}} \hat{T}^\g{_{\d\e}} \hat{T}^\e{_\c}{^\i} \nn\\
  &+& 9 \widehat{\nabla_\c \nabla_\b T}{^\h}{_{\a\g}}\, \hat{T}^\g{_\d}{^\i}\big) 
  -\frac{1}{45} \widehat{\nabla_\b T}{^\h}{_{\a\g}} \hat{R}^\g{_{\c\d}}{^\i}
  + \frac{1}{90} \widehat{\nabla_\b T}{^\h}{_{\a\g}} \widehat{\nabla_\d T}{^\g}{_\c}{^\i} 
  + \frac{1}{360}\widehat{\nabla_\b T}{^\h}{_{\a\g}} \hat{T}^\g{_{\c\e}} \hat{T}^\e{_\d}{^\i}
  -\frac{1}{40} \hat{T}{^\h}{_{\a\g}} \widehat{\nabla_\d R}{^\g}{_{\b\c}}{^\i}\nn\\
  &-& \frac{1}{80} \hat{T}{^\h}{_{\a\g}} \hat{R}^\g{_{\c\d}}{_\e} \hat{T}^\e{_\b}{^\i}
   - \frac{1}{80} \hat{T}{^\h}{_{\a\g}}\hat{T}^\g{_{\c\e}} \hat{R}^\e{_{\d\b}}{^\i} 
   +\frac{1}{720} \hat{T}{^\h}{_{\a\g}} \hat{T}^\g{_{\c\e}} \hat{T}^\e{_{\d\f}}\hat{T}^\f{_\b}{^\i}
  +\frac{1}{72} \hat{R}_{\g\a\b}{^\h} \hat{R}^\g{_{\c\d}}{^\i}
  + \h \leftrightarrow \i \Big] \nn\\
  &+& \frac{1}{9} \Big(\hat{R}_{\a(\f}{^{\h)}}{_\b} + \frac{1}{4} \hat{T}{_{\f\a\e}} \hat{T}^\e{_\b}{^\h}\Big) \Big(\hat{R}_{\c}{^{(\f\i)}}{_\d} + \frac{1}{4} \hat{T}{^\f}{_{\c\e}} \hat{T}^\e{_\d}{^\i}\Big),\label{f4}\\
  \up{(4)}{\mathcal{S}}{^{\i}}{_{\a\b\c}}&=& \up{(4)}{\mathcal{F}}{^{\i\h}}{_{\h\a\b\c}}+  \up{(4)}{\mathcal{F}}{^{\i\h}}{_{\a\h\b\c}}+  \up{(4)}{\mathcal{F}}{^{\i\h}}{_{\a\b\h\c}}+  \up{(4)}{\mathcal{F}}{^{\i\h}}{_{\a\b\c\h}}, \label{s4}\\ 
 \up{(4)}{\mathcal{P}}{_{\a\b}} &=&\frac{1}{2} \up{(2)}{\mathcal{S}}{^{\i}}{_{\a}}\up{(2)}{\mathcal{F}}{^\h}{_\h}{_{(\i\b)}} - \frac{1}{4} \up{(2)}{\mathcal{F}}{^\h}{_\h}{^{(\i}}{_{\a)}} \up{(2)}{\mathcal{F}}{^\h}{_\h}{_{(\i\b)}} +\frac{1}{2} \up{(2)}{\mathcal{F}}{^\f}{_\f}{_{\h\i}} \up{(2)}{\mathcal{F}}{^{\h\i}}{_{\a\b}} + \frac{1}{2} \big( \up{(2)}{\mathcal{F}}{^{\h\i\f}}{_\a}  \up{(2)}{\mathcal{F}}{_{\h\i(\f\b)}} + \up{(2)}{\mathcal{F}}{^{\h\i(\f}}{_{\a)}}  \up{(2)}{\mathcal{F}}{_{\h\i\b\f}}   \big) \nn\\
 &+&\frac{1}{2}  \up{(2)}{\mathcal{F}}{^{\h\i}}{_{\a\b}}  \up{(2)}{\mathcal{F}}{_{\h\i}}{^\f}{_\f}+ \frac{1}{2}\big( \up{(4)}{\mathcal{F}}{^\h}{_\h}{^\i}{_{\i\a\b}}+ \up{(4)}{\mathcal{F}}{^\h}{_\h}{^\i}{_{\a\i\b}}+ \up{(4)}{\mathcal{F}}{^\h}{_\h}{^\i}{_{\a\b\i}} + \up{(4)}{\mathcal{F}}{^\h}{_\h}{_\a}{^\i}{_{\b\i}} + \up{(4)}{\mathcal{F}}{^\h}{_\h}{_\a}{^\i}{_{\i\b}} + \up{(4)}{\mathcal{F}}{^\h}{_\h}{_{\a\b}}{^\i}{_\i}   \big). \label{p4}
\eea
\end{appendix}
\end{widetext}


%





%

%

\end{document}